\input epsf
\magnification=1200
\centerline{A REVIEW OF LINEAR RESPONSE THEORY}
\centerline{FOR GENERAL DIFFERENTIABLE DYNAMICAL SYSTEMS.}
\bigskip\bigskip
\centerline{by David Ruelle\footnote{*}{Math. Dept., Rutgers University, and 
IHES, 91440 Bures sur Yvette, France. email: ruelle@ihes.fr}.}
\bigskip\bigskip\bigskip\bigskip\noindent
	{\leftskip=2cm\rightskip=2cm\sl Abstract.  The classical theory of linear response applies to statistical mechanics close to equilibrium.  Away from equilibrium, one may describe the microscopic time evolution by a general differentiable dynamical system, identify nonequilibrium steady states (NESS), and study how these vary under perturbations of the dynamics.  Remarkably, it turns out that for uniformly hyperbolic dynamical systems (those satisfying the "chaotic hypothesis"), the linear response away from equilibrium is very similar to the linear response close to equilibrium: the Kramers-Kronig dispersion relations hold, and the fluctuation-dispersion theorem survives in a modified form (which takes into account the oscillations around the "attractor" corresponding to the NESS).  If the chaotic hypothesis does not hold, two new phenomena may arise.  The first is a violation of linear response in the sense that the NESS does not depend differentiably on parameters (but this nondifferentiability may be hard to see experimentally).  The second phenomenon is a violation of the dispersion relations: the susceptibility has singularities in the upper half complex plane.  These "acausal" singularities are actually due to "energy nonconservation": for a small periodic perturbation of the system, the amplitude of the linear response is arbitrarily large.  This means that the NESS of the dynamical system under study is not "inert" but can give energy to the outside world.  An "active" NESS of this sort is very different from an equilibrium state, and it would be interesting to see what happens for active states to the Gallavotti-Cohen fluctuation theorem.\par}
\vfill\eject
\noindent
{\bf 0. Introduction.}
\medskip
	The purpose of the present paper is to review the mathematics of linear response in the framework of the theory of differentiable dynamical systems.
\medskip
	Linear response theory deals with the way a physical system reacts to a small change in the applied forces or the control parameters.  The system starts in an equilibrium or a steady state $\rho$, and is subjected to a small perturbation $X$, which may depend on time.  In first approximation, the change $\delta\rho$ of $\rho$ is assumed to be linear in the perturbation $X$.
\medskip
	Apart from the linearity of the response $X\mapsto\delta\rho$, one can make various physical assumptions: time translation invariance, time reversibility, causality (the cause precedes the effect), energy conservation, closeness to equilibrium.  One can also find relations between the response to external perturbations and the spontaneous fluctuations of the system.  Studying the consequences of the above assumptions for bulk matter has yielded the Onsager reciprocity relations, the Kramers-Kronig dispersion relations, the Green-Kubo formula and the fluctuation-dissipation theorem.  Note that similar ideas have been used in the study of electrical circuits, in optics, and in particle scattering theory.  It is also possible to discuss higher order, {\it i.e.}, nonlinear response.
\medskip
	If the physical system in which we are interested is described by classical mechanics (with external forces, and a deterministic thermostat [17], [20]), its (microscopic) time evolution is given by an equation
$$	{dx\over dt}={\cal X}(x)\eqno{(0.1)}      $$
in phase space $M$.  We want to discuss the corresponding mathematical situation of a smooth (=differentiable) dynamical system $(f^t)$ on a compact manifold $M$.  In the case of continuous time, $(f^t)$ is called a {\it flow} and is determined by $(0.1)$ and $x(t)=f^tx(0)$, but we shall also consider the case of discrete time, where $f^n$ is the $n$-th iterate of a differentiable map $f:M\to M$.
\medskip
	The study of linear response for general smooth dynamical systems $(f^t)$ encounters a number of difficulties, and we shall obtain both positive and negative results.  In Section 1 we discuss how physical notions (like equilibrium, entropy production, etc.) and principles (like causality, energy conservation, etc.) can be related to mathematical concepts pertaining to smooth dynamics.  Then we shall analyze linear response for smooth dynamical systems in a number of different situations, both informally (Section 2), and rigorously (Sections 3, 4).
\medskip
	If we have quantum systems instead of classical systems as considered here, the theory of nonequilibrium is in part similar and in part very different (one cannot use finite systems, and one looses the smooth dynamics on a compact manifold).  We refer to V. Jak\v si\'c and coworkers (work in progress) for a comparison of classical and quantum nonequilibrium.  Note that there is a vast literature on linear response that we have not quoted.  Relevant to the approach discussed here is work by J.-P. Eckmann, C.-A. Pillet, G. Gallavotti, J.L. Lebowitz, H. Spohn, D.J. Evans, G.P. Morriss, W.G. Hoover, among others.
\bigskip\noindent
{\bf Acknowledgments.}
\medskip
	The work reported here was done in part at the ESI in Vienna, and in part at the Math Dept of McGill, Montreal.
\bigskip\noindent
{\bf 1. Smooth dynamics and physical interpretation.}
\medskip
	Let $(f^t)$ be a smooth dynamical system on the compact manifold $M$.  Interpreting $(f^t)$ as time evolution, we describe a physical state by an $(f^t)$-invariant probability measure $\rho$, and we assume that $\rho$ is ergodic\footnote{*}{{\it i.e.}, there is no nontrivial invariant decomposition $\rho=\alpha\rho_1+(1-\alpha)\rho_2$.  Supposing that the $\rho$-integral of continuous functions $A$ is given by time averages: $\rho(A)=\lim_{T\to\infty}{1\over T}\int_0^Tdt\,A(f^tx_0)$, then $\rho$ is almost certainly ergodic by Bogolyubov-Krylov theory (see for instance Jacobs [21] Section 11.3).}.  The invariance of $\rho$ expresses the {\it time translation invariance} of our physical system.  Certain simple time evolutions (completely integrable, for instance) turn out to be pathological from the physical point of view that interests us [48], and it is necessary to make some {\it chaoticity} assumption saying that the time evolution is sufficiently complicated to avoid the pathologies.  Let us be more specific.  One can show that an infinitesimal change $\delta x(0)$ in initial condition gives a later change $\delta x(t)\sim\exp\lambda t$ where $\lambda$ is called a {\it Lyapunov exponent} (if ${\rm dim}M=d$, there are $d$ Lyapunov exponents associated with an ergodic measure $\rho$).  A weak chaoticity assumption is that there is at least one Lyapunov exponent $\lambda>0$.  A strong assumption of this sort is the {\it chaotic hypothesis} of Gallavoti-Cohen [19], which says that $(f^t)$ is uniformly hyperbolic.  This will be explained in Section 3 (but we shall also consider systems that are not uniformly hyperbolic).
\medskip
	For general $(f^t)$ there is usually no invariant measure $\rho$ absolutely continuous with respect to Lebesgue on M ({\it i.e.}, such that $\rho$ has density $\rho(x)$ with respect to the Lebesgue measure $dx$ in local charts of $M$).  If there is an invariant measure $\rho$ smoothly equivalent\footnote{**}{This means that the density $\rho(x)$, and $1/\rho(x)$ are differentiable functions of $x$ in local charts.} to Lebesgue in local charts, and if it is ergodic, we say that $\rho$ is an {\it equilibrium state} (this generalizes the situation where $M$ is an "energy shell" $H(p,q)=$ constant for some Hamiltonian $H$, and $\rho$ is the corresponding normalized Liouville measure on $M$, assumed to be ergodic; $\rho$ is also known as the microcanonical ensemble).  A chaotic dynamical system $(f^t)$ typically has uncountably many ergodic measures.  Which one should one choose to describe a physical system?  A physically observed invariant state is known as a {\it nonequilibrium steady state} (NESS), and one can argue that it can be identified mathematically as an SRB probability measure, or {\it SRB state}.  The SRB states have been defined first in the uniformly hyperbolic case [46], [35], [10], and then in general [25], [26].  At this point we do not give a formal definition but state a consequence (which holds under some extra condition\footnote{***}{The extra condition is, in the discrete time case of a system generated by a diffeomorphism $f$, that $\rho$ has no vanishing Lyapunov exponent, in the continuous time case (flow) that there is only one vanishing Lyapunov exponent, corresponding to the direction of the flow.  See L.-S. Young [51] for further discussion.}): if the ergodic measure $\rho$ is SRB, there is a probability measure $\ell$ absolutely continuous with respect to Lebesgue (but in general not invariant) such that
$$ 	\rho(A)\buildrel\hbox{\sevenrm def}\over=\int\rho(dx)\,A(x)
=\lim_{T\to\infty}{1\over T}\int_0^Tdt\int\ell(dx)\,A(f^tx)\eqno{(1.1)}      $$
when $A$ is a complex continuous function on $M$ ($A$ is physically interpreted as an {\it observable}).  This says that an SRB measure $\rho$ is obtained from Lebesgue measure by a time average when the time tends to $+\infty$.  The choice of $+\infty$ (not $-\infty$) introduces a time asymmetry which will turn out to play in the mathematical theory the role played by {\it causality} as a physical principle.  Note that equilibrium states are SRB states, and general SRB states come as close to equilibrium states as is possible when there are no absolutely continuous invariant probability measures.
\medskip
	If $dx$ denotes the volume element for some Riemann metric on $M$ (or if $dx$ is smoothly equivalent to Lebesgue in local charts), we define the entropy of an absolutely continuous probability measure $\ell(dx)=\ell(x)dx$ by
$$	S(\ell)=-\int dx\,\ell(x)\log\ell(x)      $$
[In the formalism of equilibrium statistical mechanics, where $dx$ is the Liouville volume element, $S(\ell)$ is the {\it Gibbs entropy} associated with the density $\ell(\cdot)$].  Define $f^{t*}\ell$ such that $(f^{t*}\ell)(A)=\ell(A\circ f^t)$ and write $(f^{t*}\ell)(dx)=\ell_t(x)dx$, then $S(\ell_t)$ depends on $t$, and in the case of the time evolution $(0.1)$ one finds
$$	{d\over dt}S(\ell_t)=\int dx\,\ell_t(x){\rm div}{\cal X}      $$
where the divergence is taken with respect to the volume element $dx$.  One can argue that minus the above quantity is the rate at which our system gives entropy to the rest of the universe.  In other words, the {\it entropy production} $e(\ell_t)$ by our system is the expectation value of $-{\rm div}{\cal X}$, {\it i.e.}, the {\it rate of volume contraction} in $M$.  This definition extends to probability measures that are not absolutely continuous.  In particular, $(1.1)$ shows that $e(\rho)=\rho(-{\rm div}{\cal X})$ is the appropriate definition of the entropy production in the SRB state $\rho$.  Note that, since $\rho$ is invariant, $e(\rho)$ does not depend on the choice of the volume element $dx$ (contrary to $S(\ell)$).  The identification of the entropy production rate $e(\rho)$ (a physical quantity) with the expectation value of the phase space volume contraction rate (a mathematical quantity) is important\footnote {*}{The first reference we have for this identification is Andrey [1]}.  It permits in particular a study of the fluctuations of the entropy production, leading to the fluctuation theorem of Gallavotti and Cohen [19].
\medskip
	We perturb the time evolution $(f^t)$ in the continuous time case by writing
$$	{dx\over dt}={\cal X}(x)+X_t(x)      $$
instead of $(0.1)$.  In the discrete time case, the smooth map $f$ is replaced by $f+X_t\circ f$ ({\it i.e.}, $fx$ is replaced by $fx+X_t(fx)$, $t$ integer).  The perturbation $X_t$ is assumed to be infinitesimal, and may depend on the time $t$.
\medskip
	We discuss first the continuous time case, assuming that a linear response $\delta_t\rho$ is defined, and proceeding with the usual physical arguments (see Toll [49]).  We take the expectation value $\delta_t\rho(A)$ of an observable $A$ and assume for simplicity that $X_t(x)=X(x)\phi(t)$.  Linearity and time translation invariance then imply the existence of a {\it response function} $\kappa$ such that
$$	\delta_t\rho(A)=\int dt'\,\kappa(t-t')\phi(t')      $$
The Fourier transform of $\delta_t\rho(A)$ is then
$$	\int dt\,e^{i\omega t}\delta_t\rho(A)=\hat\kappa(\omega)\hat\phi(\omega)      $$
where the Fourier transform $\hat \kappa$ of the response function is called the {\it susceptibility}.  Note that, since the right-hand side is a product, there are no frequencies in the linear response that are not present in the signal $\phi$.  (Nonlinear response, by contrast, introduces harmonics and other linear combinations of the frequencies present in the signal).  If $\phi$ is square-integrable, $\int d\omega\,|\hat\phi(\omega)|^2$ may, in many physical situations, be interpreted as the energy contained in the ingoing signal.  If we assume that our system does not increase the energy in the signal ({\it conservation of energy}) we see that the susceptibility must be bounded: $|\hat\kappa(\cdot)|<{\rm constant}$.  Note that our physical assumption of "energy conservation" need not apply to a general dynamical system. In the discrete time case, the situation is analogous to that just described, Fourier transforms are replaced by Fourier series and it is convenient to introduce the variable $\lambda=e^{i\omega}$ so that the susceptibility is replaced by a function $\Psi(\lambda)$.
\medskip
	In our physical discussion, {\it causality} is expressed by the fact that $\kappa(t-t')$ vanishes when $t<t'$.  This, together with the boundedness of the susceptibility, implies that $\hat\kappa(\cdot)$ extends to a bounded analytic function in the upper half complex plane, and that the real and imaginary parts of $\hat\kappa(\omega)$ (for $\omega\in{\bf R}$) satisfy integral relations known as the Kramers-Kronig {\it dispersion relations} (see the discussion in [49]).  When dealing with a general dynamical system, causality is replaced by the assumption that the state $\rho$ of our system is an SRB state.  To clarify this point we shall in the next section make a nonrigorous calculation of $\delta_t\rho$, and see how causality appears.  We shall also discuss the special case when $\rho$ is an equilibrium state (perturbation {\it close to equilibrium}) and understand how the {\it fluctuation-dissipation} relation arises.  Away from equilibrium only part of the fluctuation-dissipation relation will survive.
\bigskip\noindent
{\bf 2. Linear response: an informal discussion.}
\medskip
	We shall now evaluate the linear response $\delta\rho$ by a nonrigorous calculation (using what is called first order perturbation theory in physics).  Le us consider a discrete dynamical system $(f^n)$ where $f:M\to M$ is a smooth map, and the formula $(1.1)$ is replaced by
$$	\rho(A)=\lim_{n\to\infty}{1\over n}\sum_{k=1}^n\int\ell(dx)\,A(f^kx)
	=\lim_{n\to\infty}{1\over n}\sum_{k=1}^n\int(f^{k*}\ell)(dx)\,A(x)\eqno{(2.1)}      $$
In $(2.1)$, we assume that $\ell(dx)=\ell(x)dx$ is an absolutely continuous probability measure on $M$.  The formula $(2.1)$ holds if $\rho$ is an SRB measure for a diffeomorphism $f$ of $M$, and also if $\rho$ is an absolutely continuous invariant measure (a.c.i.m.) for a map $f$ of an interval $[a,b]\subset{\bf R}$.  Replacing $fx$ by $\tilde fx=fx+X(fx)$ we have to first order in $X$
$$	\tilde f^kx=f^kx+\sum_{j=1}^k(T_{f^jx}f^{k-j})X(f^jx)      $$
where $T_xf$ denotes the tangent map to $f$ at the point $x$.  Thus
$$	A(\tilde f^kx)=A(f^kx)+A'(f^kx)\sum_{j=1}^k(T_{f^jx}f^{k-j})X(f^jx)      $$
$$	=A(f^kx)+\sum_{j=1}^kX(f^jx)\cdot\nabla_{f^jx}(A\circ f^{k-j})      $$
hence
$$	\delta\rho(A)=\lim_{n\to\infty}{1\over n}
	\sum_{k=1}^n\sum_{j=1}^k\int\ell(dx)\,X(f^jx)\cdot\nabla_{f^jx}(A\circ f^{k-j})      $$
$$	=\lim_{n\to\infty}{1\over n}\sum_{k=1}^n\sum_{j=1}^k
	\int(f^{j*}\ell(dx))\,X(x)\cdot\nabla_x(A\circ f^{k-j})      $$
$$	=\lim_{n\to\infty}{1\over n}\sum_{i\ge0}\sum_{j=1}^{n-i}
	\int(f^{j*}\ell(dx))\,X(x)\cdot\nabla_x(A\circ f^i)      $$
If we interchange in the right-hand side $\lim_{n\to\infty}$ and $\sum_{i\ge0}$ (without a good mathematical justification!), and use $\lim_{n\to\infty}{1\over n}\sum_{j=1}^{n-i}f^{j*}\ell=\rho$ we obtain formally
$$	\delta\rho(A)=\sum_{n=0}^\infty\int\rho(dx)\,X(x)\cdot\nabla_x(A\circ f^n)
=\sum_{n=0}^\infty\int\rho(dx)\,X(f^{-n}x)\cdot\nabla_{f^{-n}x}(A\circ f^n)\eqno{(2.2)}$$
The physical meaning of this formula is that the change of $\rho(A)$ due to the perturbation $X$ is a sum over $n$ of terms corresponding to the perturbation acting at time $-n$.  The fact that the sum extends over $n\ge0$ may be interpreted as {\it causality}, and results from the asymmetry in time of the formula $(2.1)$ defining the SRB measure $\rho$.  Replacing $X(f^{-n}x)$ by $e^{in\omega}X(f^{-n}x)$ in the right-hand side of $(2.2)$ we obtain the susceptibility which, as a function of $\lambda=e^{in\omega}$, is
$$	\Psi(\lambda)
=\sum_{n=0}^\infty\lambda^n\int\rho(dx)\,X(x)\cdot\nabla_x(A\circ f^n)\eqno{(2.3)}      $$
Since $f$ has bounded derivatives on $M$, the power series in $\lambda$ defined by the right-hand side of $(2.3)$ has nonzero radius of convergence.  Formally, $\delta\rho(A)=\Psi(1)$, but we are not assured that $\Psi(1)$ makes sense.
\medskip
	In the continuous time case $(2.3)$ is replaced by
$$	\hat\kappa(\omega)
=\int_0^\infty dt\,e^{i\omega t}\int\rho(dx)\,X(x)\cdot\nabla_x(A\circ f^t)\eqno{(2.4)}      $$
and formally $\delta\rho(A)=\hat\kappa(0)$ (this corresponds to taking $X_t(x)=X(x)$ or $\phi (t)=1$ in Section 1) but we are not assured that $\hat\kappa(0)$ makes sense.  Comparing with the physical discussion of Section 1, we note that $\kappa(t)=0$ for $t<0$, {\it i.e.}, causality is satisfied.  But it may happen (see Section 4) that $\kappa(t)=\int\rho(dx)\,X(x)\cdot\nabla_x(A\circ f^t)$ grows exponentially with $t$, so that $\hat\kappa(\omega)$ does not extend analytically to the upper half plane.  This apparent "violation of causality" in fact means that "conservation of energy" is violated: when hit by the periodic perturbation $e^{i\omega t}X$, the system may give out (much) more energy than it receives.
\medskip
	We consider now the situation where $\rho$ is an {\it equilibrium state}.  Thus $\rho(dx)=\rho(x)dx$ in a local chart and we may define ${\rm div}_\rho X$ by
$$	{\rm div}_\rho X(x)={1\over\rho(x)}\sum_{i=1}^d{\partial\over\partial x_i}(\rho X_i)  $$
It is convenient to write simply $dx$ for the volume element $\rho(x)dx$ and ${\rm div}_x X$ for ${\rm div}_\rho X(x)$.  We obtain then
$$	\int\rho(dx)\,X(x)\cdot\nabla_x(A\circ f^n)=-\int dx\,({\rm div}_x X)A(f^nx)      $$
Notice that the right-hand side is a correlation function in the time variable $n$.  If we assume that this correlation function tends exponentially\footnote{*}{Note that $\int\rho(dx)\,{\rm div_x}X(x)=0$ so that the correlation function tends to 0 when $n\to\infty$ if the time evolution is mixing.} to 0 when $n\to\infty$ it follows that the radius of convergence of $\Psi(\lambda)$ is $>1$, and $(2.3)$ becomes
$$	\Psi(\lambda)=-\sum_{n=0}^\infty\lambda^n\int dx\,({\rm div}_xX)A(f^nx)      $$
which makes sense for $\lambda=1$. Similarly we obtain from $(2.4)$ in the continuous time case:
$$	\hat\kappa(\omega)
=-\int_0^\infty dt\,e^{i\omega t}\int dx\,({\rm div}_x X)A(f^tx)\eqno{(2.5)}      $$
The susceptibility $\hat\kappa(\omega)$ appearing in the left-hand side of $(2.5)$ gives the linear response of our dynamical system to a periodic signal which puts the system outside of equilibrium, {\it i.e.}, in a so-called {\it dissipative} regime.  The right-hand side is constructed from a time correlation function which describes the {\it fluctuations} of our system in the equilibrium state.  The relation between dissipation and fluctuations expressed by $(2.5)$ is a form of the so-called {\it fluctuation-dissipation theorem} ($(2.5)$ is also related to the Green-Kubo formula).  A physical interpretation of the fluctuation-dissipation theorem is that kicking the system outside of equilibrium by the perturbation $X$ is equivalent to waiting for a spontaneous fluctuation that has the same effect as the kick.  The reason that this is possible is the absolute continuity of $\rho$.
\medskip
	If we assume that the correlation function in the right-hand side tends exponentially\footnote{*}{Exponential decay of correlations does not always hold for smooth dynamical systems, even if they are uniformly hyperbolic, but is still a natural assumption [15].  In the statistical mechanics of bulk matter, time correlation functions decay more slowly, say like $t^{-\nu/2}$ which holds for diffusion in $\nu$ dimensions, and it is natural to assume absolute integrability in time, so that the susceptibility is analytic only in the upper half plane.  Incidentally, this means that bulk matter with dimension $\nu<3$ is expected to behave pathologically with respect to linear response.} to 0 when $t\to\infty$, we see that $\hat\kappa(\cdot)$ extends to an analytic function in $\{\omega\in{\bf C}:{\rm Im}\omega>-\epsilon\}$ for some $\epsilon>0$ and that the real and imaginary parts of $\hat\kappa(\cdot)$ on ${\bf R}$ are related by Hilbert transforms: these relations are the {\it Kramers-Kronig dispersion relations} (see Toll [49]).
\medskip
        We return now to the study of the susceptibility (2.3) or (2.4) when $\rho$ is an SRB but not necessarily an equilibrium state.  At this point we need a brief description of the ergodic theory of smooth dynamical systems following the ideas of Oseledec (see [31], [36]), Pesin [33], [34], Ledrappier, Strelcyn, and Young (see [25], [26]).  For definiteness we discuss the discrete time case of a diffeomorphism $f:M\to M$ where $M$ has dimension $d$.  Given an ergodic measure $\rho$ for $f$, there are $d$ Lyapunov exponents $\lambda_1\le\ldots\le\lambda_d$ which give the possible rates of exponential separation of nearby orbits (almost everywhere with respect to $\rho$).  For $\rho$-almost every $x$, there are a {\it stable} (or contracting) smooth manifold ${\cal V}_x^s$ and an {\it unstable} (or expanding) smooth manifold  ${\cal V}_x^u$ through $x$.  The dimension of ${\cal V}_x^s$ is the number of Lyapunov exponents $<0$, and the dimension of ${\cal V}_x^u$ is the number of Lyapunov exponents $>0$.  The manifold ${\cal V}_x^s$ is shrunk exponentially under iterates of $f$ while ${\cal V}_x^u$ is shrunk exponentially under iterates of $f^{-1}$.  The ergodic measure $\rho$ is SRB if and only if it is {\it smooth along unstable directions}.  This may be taken as a general definition of an SRB measure, and means that there is a set $S$ with $\rho(S)=1$, and a partition of $S$ into pieces $\Sigma_\alpha\subset{\cal V}_\alpha^u$ of unstable manifolds such that the conditional measure $\sigma_\alpha$ of $\rho$ on $\Sigma_\alpha$ is absolutely continuous with respect to Lebesgue on ${\cal V}_\alpha^u$.  The manifolds ${\cal V}_x^s,{\cal V}_x^u$ do not depend continuously on $x$ (only measurably).  If ${\cal V}_x^u$ and ${\cal V}_y^u$ are unstable manifolds with $y$ close to $x$, one can define a holonomy map $\pi$ of part of ${\cal V}_y^u$ to part of ${\cal V}_x^u$ along the stable manifolds (a stable manifolds through a point of ${\cal V}_y^u$ may ``curve back'' before it 
 hits ${\cal V}_x^u$ but, in terms of Lebesgue measure, most of ${\cal V}_y^u$ is mapped to most of ${\cal V}_x^u$ if $y$ is sufficiently close to $x$), and the map $\pi$ is absolutely continuous\footnote {**}{This is a form of Pesin's theorem of absolute continuity of foliations.  See for instance [6] p.302.}: sets of Lebesgue measure 0 are sent to sets of Lebesgue measure 0.
\medskip
        Suppose now that the SRB measure $\rho$ for $f$ has no vanishing Lyapunov exponent.  We can then write (for $\rho$-almost all $x$) $X(x)=X^s(x)+X^u(x)$ where $X^s(x)$ is in the stable direction (tangent to ${\cal V}_x^s$) while $X^u(x)$ is in the unstable direction (tangent to ${\cal V}_x^u$).  Inserting this in (2.3) we find
$$      \Psi(\lambda)
=\sum_{n=0}^\infty\lambda^n\int\rho(dx)X^s(x)\cdot\nabla_x(A\circ f^n)
+\sum_{n=0}^\infty\lambda^n\int\rho(dx)X^u(x)\cdot\nabla_x(A\circ f^n)     $$
The $X^s$-integral may be rewritten as
$$  \int\rho(dx)((T_{f^{-n}x}f^n)X^s(f^{-n}x))\cdot\nabla_xA      $$
where $(Tf^n)X^s$ decreases exponentially with $n$.  Since $\rho$ is an SRB measure, the $X^u$-integral may be rewritten in terms of integrals with respect to $\sigma_\alpha(dx)$ on pieces $\Sigma_\alpha$ of unstable manifolds ${\cal V}_\alpha^u$, where $\sigma_\alpha(dx)$ is absolutely continuous with respect to Lebesgue on ${\cal V}_\alpha^u$.  Introducing a divergence ${\rm div}^u$ in the unstable direction one may hope to rewrite the $X^u$-integral as
$$    -\int\rho(dx)({\rm div}_{\sigma_\alpha}^uX^u(x))A(f^nx)      $$
This is a correlation function with respect to the time variable $n$, and one may hope that it tends to 0 when $n\to\infty$.  In conclusion one may hope that $\Psi(\lambda)$ has radius of convergence $>1$ and that $\delta\rho(A)=\Psi(1)$.
\medskip
        In the continuous time case, the hope is that we may rewrite (2.4) as
$$      \hat\kappa(\omega)=\hat\kappa^s(\omega)+\hat\kappa^{cu}(\omega)      $$
$$      \hat\kappa^s(\omega)=\int_0^\infty dt\,e^{i\omega t}
\int\rho(dx)((T_{f^{-t}x}f^t)X^s(f^{-t}x))\cdot\nabla_xA      $$
$$      \hat\kappa^{cu}(\omega)=-\int_0^\infty dt\,e^{i\omega t}
\int\rho(dx)({\rm div}_{\sigma_\alpha}^{cu}X^{cu}(x))A(f^tx)      $$
where $X^{cu}$ is the component of $X$ in the center-unstable direction corresponding to the Lyapunov exponents $\ge0$ (one zero exponent for the ``flow direction'', {\it i.e.}, the direction of ${\cal X}$).  An optimistic guess would be that $\hat\kappa(\omega)$ extends to a holomorphic function for ${\rm Im}\omega>-\epsilon$ with $\epsilon>0$, and $\delta\rho(A)=\hat\kappa(0)$.  Note that $\hat\kappa^{cu}$ is formally the Fourier transform of a time correlation function (cut to $t\ge0$), {\it i.e.}, $\hat\kappa^{cu}$ formally conforms to the fluctuation-dissipation theorem (as in the case where $\rho$ is an equilibrium state).  In order to interpret $\hat\kappa^s$ remember that, if $\rho$ is not an equilibrium state, $\rho$ is singular with respect to Lebesgue.  One might say that $\rho$ is concentrated on an attractor\footnote{*}{The idea that $\rho$ is concentrated on an attractor $\subset M$ is geometrically appealing, but in fact the support of $\rho$ may be the whole of 
 $M$.  The geometric notion of an attractor in $M$ should thus be replaced by the idea of $\rho$ as a measure-theoretic attracting point.} $\ne M$.  A perturbation that kicks the system in the stable direction away from the attractor is not equivalent to a spontaneous fluctuation.  The effect of such a perturbation (the system oscillates and tends to the attractor) is described by $\hat\kappa^s$ [39], [28].
\medskip
        In Section 3 we shall see that in the uniformly hyperbolic case things work out basically as indicated above.  In general however, $\rho(A)$ is not expected to depend differentiably on the dynamical system.  In order to understand what happens to linear response in nonhyperbolic situations we shall discuss in Section 4 the case of unimodal maps $f$ of the interval and see how $\rho(A)$ may depend nondifferentiably on $f$, yet look differentiable for the purposes of physics.
\bigskip\noindent
{\bf 3. Linear response: the uniformly hyperbolic case.}
\medskip
        In this Section we shall discuss the case of a dynamical system $(f^t)$ restricted to a neighborhood $U$ of a {\it hyperbolic attractor}\footnote{*}{Hyperbolic attractors are also called Axiom A attractors [47].  Hyperbolicity as defined here is uniform hyperbolicity.  Weaker forms of hyperbolicity are not considered in this Section.} $K\subset M$.  (This includes the situation that $f$ is an Anosov diffeomorphism of $M$ or $(f^t)$ an Anosov flow on $M$, in those situations $K=U=M$).  For completeness we now give a certain number of definitions.  These definitions make use of a Riemann metric on $M$, but do not depend on the choice of the metric.
\medskip
        If $f$ is a diffeomorphism, we say that $K$ is a {\it hyperbolic set for $f$} if $T_KM$ (the tangent bundle restricted to $K$) has a continuous $Tf$-invariant splitting $T_KM=E^s\oplus E^u$, and there are constants $c,\lambda>0$ such that
$$     ||Tf^nv||\le ce^{-n\lambda}||v||\qquad{\rm if}\qquad v\in E^s,n\ge0    $$
$$   ||Tf^{-n}v||\le ce^{-n\lambda}||v||\qquad{\rm if}\qquad v\in E^u,n\ge0    $$
($E^s$ and $E^u$ are called the stable and unstable subbundles of $T_KM$).
\medskip
       If $(f^t)$ is the flow associated with a vector field ${\cal X}$ (see (0.1)), we assume (for simplicity) that ${\cal X}$ does not vanish on $K$ and we say that $K$ is a {\it hyperbolic set for $(f^t)$} if $T_KM$ has a continuous invariant splitting
 $T_KM=E^c\oplus E^s\oplus E^u$, and there are constants $c,\lambda>0$ such that
$$       E^c\hbox{ is one-dimensional and }E_x^c={\bf R}{\cal X}(x)      $$
$$     ||Tf^tv||\le ce^{-\lambda t}||v||\qquad{\rm if}\qquad v\in E^s,t\ge0    $$
$$   ||Tf^{-t}v||\le ce^{-\lambda t}||v||\qquad{\rm if}\qquad v\in E^u,t\ge0   $$
\medskip
        We say that $K$ is a {\it basic hyperbolic set} (for a diffeomorphism or flow) if

        (a) $K$ is a compact invariant set as above,

        (b) the periodic orbits of $f^t|K$ are dense in $K$,

        (c) $f^t|K$ is topologically transitive, {\it i.e.}, the orbit $(f^tx)$ of some $x\in K$ is dense in $K$,

        (d) the open set $U\supset K$ can be chosen such that $\cap_tf^tU=K$.
\medskip
        In particular we say that $K$ is a {\it hyperbolic attractor} (or Axiom A attractor in the terminology of Smale [47]) if one can chose $U$ such that $f^tU\subset U$ for all sufficiently large $t$.  We have then $K=\cap_{t\ge0}f^tU$.
\medskip
        Uniform hyperbolicity has been much studied in relation with {\it structural stability}: to a small perturbation of $(f^t)$ corresponds a small perturbation of the hyperbolic set $K$.  In particular, if the diffeomorphism $f$ has a hyperbolic attractor $K$, and $\tilde f$ is close to $f$ in a suitable C$^r$ topology, then $\tilde f$ has a hyperbolic attractor $\tilde K$ and there is a homeomorphism $h:K\to\tilde K$ close to the identity such that $\tilde f|\tilde K=h\circ f\circ h^{-1}$.  In the case of a flow $(f^t)$ there is a similar result, but a reparametrization of $\tilde f^t|\tilde K$ is necessary.  An important tool in the study of uniformly hyperbolic dynamical systems is constituted by {\it Markov partitions} (introduced by Sinai [44], [45], with important contributions by Bowen [7]. [8], [9]).  Using Markov partitions, it is possible to replace problems about measures on a hyperbolic attractor by problems of equilibrium statistical mechanics of one-dimensional spin systems.  The transition is via {\it symbolic dynamics}, and makes available some efficient tools like {\it transfer operators} (introduced by Ruelle, see in particular [2] which is a good source of references on earlier literature).  The body of knowledge accumulated in this direction is known as {\it thermodynamic formalism} (see [37], [32]).
\medskip
        Of interest to us here are the results concerning SRB measures.  On a hyperbolic attractor $K$ there is exactly one SRB measure $\rho$.  Writing $\rho(A)=\int\rho(dx)A(x)$ we may characterize $\rho$ as follows (with $U\supset K$ as above).

        (i) For Lebesgue-almost every $x\in U$,
$$      \rho(A)=\lim_{N\to\infty}{1\over N}\sum_{n=0}^{N-1}A(f^nx)
\qquad\hbox{(diffeomorphism case)}      $$
$$      \rho(A)=\lim_{T\to\infty}{1\over T}\int_0^Tdt\,A(f^tx)
\qquad\hbox{(flow case)}      $$
if $A$ is a continuous function $U\to{\bf C}$.

        (ii) If $\ell(x)dx$ is a probability measure with support in $U$ and absolutely continuous with respect to Lebesgue, then
$$ \rho(A)=\lim_{N\to\infty}{1\over N}\sum_{n=0}^{N-1}\int\ell(x)dx\,A(f^nx)  $$
in the diffeomorphism case, and similarly in the flow case.
\medskip
        There is a rich geometric theory of hyperbolic sets (beginning with [47])  which we cannot properly describe here.  Let us mention that for every point $x$ of a hyperbolic set there are a stable and an unstable manifold ${\cal V}_x^s$, ${\cal V}_x^u$, exponentially contracted by $f^t$ or $f^{-t}$ when $t\to\infty$.  If $K$ is a hyperbolic attractor, then ${\cal V}_x^u\subset K$.  In the flow case, it is useful to consider the center-unstable manifold ${\cal V}_x^{cu}$, union over $t$ of the $f^t{\cal V}_x^u$.  The SRB measure $\rho$ is the only ergodic measure on $K$ such that its conditional measures on unstable (or center-unstable) manifolds are absolutely continuous with respect to Lebesgue on those manifolds (see Section 2).
\medskip
        There are other important characterizations of the SRB measure $\rho$ that will, however, not be used here.  One of them is that the invariant\footnote{*}{See [5].  The Kolmogorov-Sinai invariant $h_{KS}$ is also known as {\it entropy}, but should not be confused with the Gibbs entropy discussed in Section 1.} $h_{KS}(\rho)$ (rate of creation of information in the state $\rho$) is equal to the sum of the positive Lyapunov exponents of $\rho$ (this sum is also equal to the expectation value in $\rho$ of the logarithm of the unstable Jacobian = rate of growth of the unstable volume element).  Another is that $\rho$ is stable under small stochastic perturbations of the dynamics.
\medskip
        We now have the concepts that allow us to discuss the dependence of the SRB measure $\rho$ on the dynamical system $(f^t)$.  We first discuss the discrete time case.
\medskip
        {\sl Let $K_0$ be a hyperbolic ({\it i.e.}, Axiom A) attractor for the {\rm C}$^3$ diffeomorphism $f_0$, and suppose for simplicity that $f_0|K_0$ is mixing\footnote{**}{\rm This is equivalent to requiring that $K$ is connected.  In general $K$ would consist of $m$ connected components permuted by $f$, and one reduces to the mixing situation by considering $f^m$ restricted to one of the connected components.}.  If $f$ is allowed to vary in a small neighborhood of $f_0$, there is a hyperbolic attractor $K$ for $f$, depending continuously on $f$ and a unique SRB measure $\rho$ for $f$ with support $K$.  Furthermore
\medskip
        (a) there is a {\rm C}$^3$ neighborhood ${\cal N}$ of $f_0$ such that if $A:M\to{\bf R}$ is {\rm C}$^2$, then $f\mapsto\rho(A)$ is differentiable in ${\cal N}$,
\medskip
        (b) the first-order change $\delta\rho(A)$ when $f$ is replaced by $f+X\circ f$ is given by $\delta\rho(A)=\Psi(1)$, where the power series
$$      \Psi(\lambda)
=\sum_{n=0}^\infty\lambda^n\int\rho(dx)X(x)\cdot\nabla_x(A\circ f^n)      $$
has a radius of convergence $>1$,
\medskip
        (c) if $X^s,X^u$ are the components of $X$ in the stable and unstable directions, we may write $\Psi(\lambda)=\Psi^s(\lambda)+\Psi^u(\lambda)$, where the power series
$$      \Psi^s(\lambda)
=\sum_{n=0}^\infty\lambda^n\int\rho(dx)((T_xf^n)X^s)\cdot\nabla_{f^nx}A      $$
$$      \Psi^u(\lambda)
=-\sum_{n=0}^\infty\lambda^n\int\rho(dx)({\rm div}_x^uX^u)A(f^nx)      $$
both have radius of convergence $>1$.}
\medskip
        The proof of (a) appears in [24], while (a),(b),(c) are proved in [38].
\medskip
        The divergence in the unstable direction ${\rm div}_x^uX^u$ that appears above can be shown to be a H\"older continuous function.  Therefore, the coefficients of the power series $\Psi^u$ are the values of a correlation function $n\mapsto\int\rho(x)A(f^nx)B(x)$ tending exponentially to 0.  One can also show that $\Psi^u(\lambda)$ extends meromorphically to a circle $|\lambda|<R$ with $R>1$; its poles are Ruelle-Pollicott {\it resonances} (see [2])corresponding to fluctuations of the system $((f^t),\rho)$ in accordance with the fluctuation-dissipation theorem.  The poles of $\Psi^s(\lambda)$ would correspond to resonances in the ``oscillations of the system around its attractor''.  Near equilibrium, {\it i.e.}, when $\rho$ is absolutely continuous and we may write $\rho(dx)=dx$, we find that the coefficients of $\Psi^s$ are also the values of a correlation function, and the two kinds of poles become the same.  It would be interesting to discuss examples where the ``stable'' resonances separate from the ``unstable'' resonances as one moves away from equilibrium.
\medskip
        The following figure shows the singularities of $\Psi(\lambda)$ in the complex $\lambda$-plane: no pole for $|\lambda|\le1$, some poles (crosses) for $1<|\lambda|<R$, possible essential singularities for $|\lambda|\ge R$.
\medskip
\centerline{\epsfbox{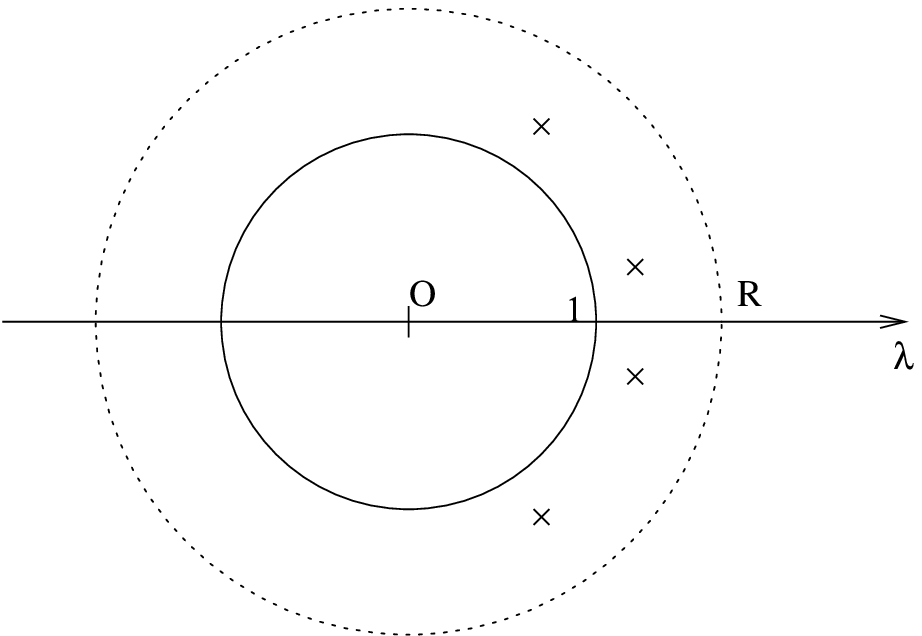}}
\medskip
        We discuss now the flow case (continuous time).
\medskip
        {\sl Let the ${\rm C}^3$ vector field ${\cal X}+aX$ on $M$ define a flow $(f_a^t)$ with a hyperbolic ({\it i.e.}, Axiom A) attractor $K_a$ depending continuously on $a\in(-\epsilon,\epsilon)$.  There is then a unique SRB measure $\rho_a$ for $(f_a^t)$ with support $K_a$.  Furthermore
\medskip
        (a) if $A:M\to{\bf R}$ is ${\rm C}^2$, then $a\mapsto\rho_a(A)$ is ${\rm C}^1$ on $(-\epsilon,\epsilon)$,
\medskip
        (b) the derivative $d\rho_a(A)/da$ is the limit when $\omega\to0$ for $\omega>0$ of 
$$      \hat\kappa_a(\omega)
=\int_0^\infty e^{i\omega t}dt\int\rho_a(dx)X(x)\cdot\nabla_x(A\circ f_a^t)   $$
where $\hat\kappa_a(\omega)$ is holomorphic for ${\rm Im}\omega>0$,
\medskip
        (c) the function $\omega\mapsto\hat\kappa_a(\omega)$ extends meromorphically to $\{\omega:{\rm Im}\omega>-\Lambda\}$, and the extension has no pole at $\omega=0$.}
\medskip
        This is proved in [42], using the machinery of Markov partitions.  A different proof has been given in [11] in the case of Anosov flows.
\medskip
        Note that we did not require $(f_a^t)$ to be mixing, and $\hat\kappa_a(\omega)$ may thus have poles on the real axis.  If $(f_a^t)$ is mixing, $\hat\kappa_a(\omega)$ might still have poles arbitrarily close to the real axis.  To obtain a discussion analogous to that for diffeomorphisms, it is natural to define a center-unstable divergence $C={\rm div}^{cu}(X^c+X^u)$, where we take $a=0$ for simplicity.  If $\int dt|\rho_0((A\circ f_0^t)C)|<\infty$ we have
$$      {d\over da}\rho_a(A)|_{a=0}
=\int_0^\infty dt\int\rho_0(dx)X(x)\cdot\nabla_x(A\circ f_0^t)      $$
If the correlation function $t\mapsto\rho_0((A\circ f_0^t)C)$ tends to 0 exponentially at $\infty$, the poles of $\hat\kappa_0(\omega)$ stay a finite distance away from the real axis.  A number of results are known on the decay of correlation functions for hyperbolic flows (see in particular Chernov [13], Dolgopyat [14], [15], Liverani [27], Fields {\it et al.} [18]).
\medskip
        The following figure shows the singularities of $\hat\kappa(\omega)$ in the complex $\omega$-plane: no poles for ${\rm Im}\omega>0$, some poles for $0\ge{\rm Im}\omega>-\Lambda$, possible essential singularities for ${\rm Im}\omega\le-\Lambda$
\medskip
\centerline{\epsfbox{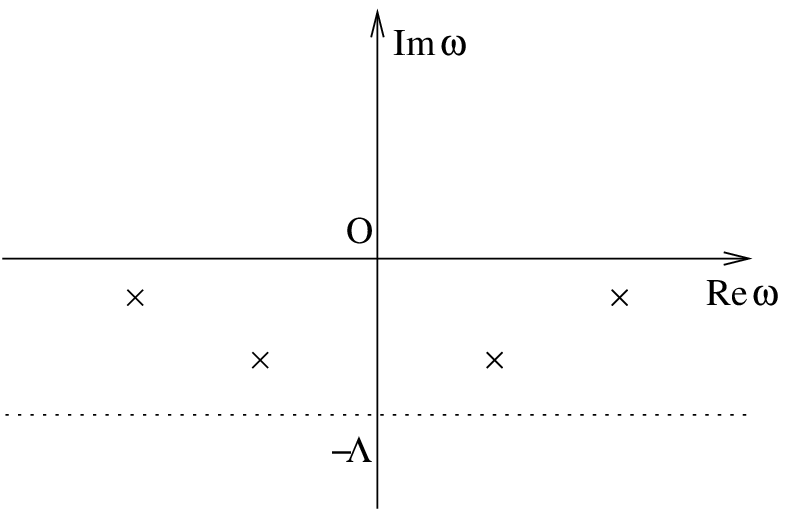}}
\medskip\noindent
To compare with diffeomorphisms, remember that $\lambda=e^{i\omega}$.
\medskip
	Let us mention at this point a class of dynamical systems that are "almost" uniformly hyperbolic, namely the Lorenz system and related flows (see for instance Chapter 9 in [6]).  Numerical studies [29] seem to indicate that the Lorenz system behaves like uniformly hyperbolic dynamical systems with respect to linear response.  A mathematical study would here be very desirable.  For partially hyperbolic systems see [16].
	
\bigskip\noindent
{\bf 4. Linear response: the case of unimodal maps.}
\medskip
        Among dynamical systems that are not uniformly hyperbolic, the unimodal maps of the interval have been particularly well studied.  These are smooth noninvertible maps $f:I\to I$, where $I=[a,b]$ is a compact interval in ${\bf R}$.  One assumes that $a<c<b$ and that $f'(x)>0$ for $x\in[a,c)$, $f'(x)<0$ for $x\in(c,b]$: this makes $f$ one-humped (unimodal).  It is known since Jakobson [22] that many unimodal maps have an a.c.i.m. $\rho(x)dx$, {\it i.e.}, an invariant measure absolutely continuous with respect to Lebesgue on $[a,b]$.  (The concept of an SRB measure discussed earlier is replaced by that of an a.c.i.m. $\rho(dx)=\rho(x)dx$ in the present 1-dimensional situation).  Consider the specific example of maps $f_k:[0,1]\to[0,1]$ defined by $f_kx=kx(1-x)$.  These have an a.c.i.m. $\rho_k$ for a set $S$ of values of $k$ with positive Lebesgue measure in $(0,4]$.  But one also knows that the complement of $S$ is dense in $(0,4]$.  How then could the function $k\to\rho_k$ be differentiable?  One idea [40] is to use differentiability in the sense of Whitney [50]: find a differentiable function $\phi$ such that $\phi(k)=\rho_k(A)$ when $k$ belongs to some set $\Sigma\subset S$ and define the derivative $d\rho_k(A)/dk$ to be $\phi'(k)$ when $k\in\Sigma$.
\medskip
        The above considerations suggest the following program: start with a unimodal map $f$ with an a.c.i.m. $\rho$, perturb $f$ to $f+X\circ f$ and define the corresponding derivative of $\rho(A)$ to be
$$      \delta\rho(A)=\Psi(1)      $$
where
$$      \Psi(\lambda)
=\sum_{n=0}^\infty\lambda^n\int\rho(dx)X(x){d\over dx}A(f^nx)      $$
Then, show that $\delta\rho(A)$ is indeed a Whitney derivative in some sense.  The conclusions of this program are as follows: {\sl

         (i) even with the idea of Whitney derivative, it appears that $f\mapsto\rho$ is (mildly) nondifferentiable,

         (ii) the radius of convergence of $\Psi$ is $<1$, {\it i.e.}, the susceptibility function $\omega\mapsto\Psi(e^{i\omega})$ has singularities in the upper half-plane.}
\medskip
        We now give an idea of how such conclusions can be reached, referring to [41], [23], [43]\footnote{*}{Let us note that Baladi and Smania [3] have made a detailed study of linear response for piecewise expanding maps of the interval.  The theory of these maps is in some respects very similar to that of smooth unimodal maps, in other respects quite different ($\Psi$ has no singularities inside the unit circle).} for the detailed assumptions and proofs.  We note first that it is no loss of generality to suppose that the interval $[a,b]$ and the critical point $c$ satisfy $fc=b,fb=a$, and we assume $a<fa<b$. If the density $\rho(\cdot)$ of the a.c.i.m. is differentiable and nonzero at $c$, the invariance of $\rho(x)dx$ under $f$ implies that $\rho(x)$ has a {\it spike} $\sim(b-x)^{-1/2}$ near $b$.  In fact, for each point $f^nb=f^{n+1}c$ of the critical orbit (with $n=0,1,\ldots$) there is a spike on one side of $f^nb$, with singularity 
$$      C_n|x-f^nb|^{-1/2}      $$
where
$$      C_n=\rho(c)|{1\over2}f''(c)\prod_{k=0}^{n-1}f'(f^kb)|^{-1/2}      $$
\indent
        The easiest situation to analyze is when the critical point is preperiodic.  Specifically we assume that $f^kc$ belongs to an unstable periodic orbit of period $\ell$ for some $k,\ell$.  One can then prove that the radius of convergence of $\Psi(\lambda)$ is $<1$.  In fact $\Psi$ has $\ell$ poles regularly spaced on a circle of radius $<1$ (but no singularity at $\lambda=1$).  The situation is thus analogous to that of a hyperbolic attractor, but with some poles inside the unit circle.  This gives an easy example of our assertion (ii) above.
\medskip
        We consider now a more general class of unimodal maps.  First, we assume that $f$ has a compact invariant hyperbolic set $K$.  This is no big deal: we can assume that $|f'(x)|>1$ for $x$ away from some neighborhood of $c$, so that the points with orbits avoiding a suitable neighborhood of $c$ automatically form a hyperbolic set $K$.  Misiurewicz [30] has shown that if some point of the critical orbit is in $K$ (say $f^3c\in K$), then $f$ has an a.c.i.m.  Here the singularities of $\Psi$ inside the unit circle are expected to be worse then poles, probably forming a natural boundary.  The interest of the Misiurewicz situation is that it can be perturbed: replacement of a Misiurewicz map $f_0$ (with hyperbolic set $K_0$, and critical point $c_0$) by $f$ replaces $K_0$ by $K$, with a homeomorphism $h:K_0\to K$ such that $f\circ h=h\circ f_0$ on $K_0$.  If $f^3c\in K$, then $f$ is again a Misiurewicz map, with an a.c.i.m. $\rho$, and we can study the dependence of $\rho$ on $f$.  The evidence is that $\rho$ does not depend differentiably on $f$.  This is because the $n$-th spike has an amplitude decreasing exponentially like $|\prod_{k=0}^{n-1}f'(f^kb)|^{-1/2}$, but it moves around at a speed that can be estimated\footnote{*}{This estimate fails if certain cancellations occur, and cancellations indeed occur when $f^3c=hf_0^3c_0$.  Those $f$ such that $f^3c=hf_0^3c_0$ form the topological conjugacy class of $f_0$, and the map $f\mapsto\rho$ can be shown to be differentiable on a topological conjugacy class [43].} to increase exponentially like $|\prod_{k=0}^{n-1}f'(f^kb)|$.
\medskip
        Note that this evidence of nondifferentiability is not a proof!  The nondifferentiability of $f\mapsto\rho$ will probably not be very important in physical situations, because the spikes of high order (which lead to nondifferentiability) will be drowned in noise and therefore invisible.  Using Misiurewicz maps we have thus argued [43] that, for such maps, $f\mapsto\rho$ is mildly nondifferentiable, in agreement with our assertion (i) above.  A study of the more general Collet-Eckmann maps is under way [4].
\medskip
        To summarize, the study of unimodal maps reveals two new phenomena that should be present in more general dynamical systems: (i)  nondifferentiability of $f\mapsto\rho$, and (ii) the apparently ``acausal'' singularities of the susceptibility in the upper half complex plane.  Of these phenomena, (ii) may be most easy to observe (see for instance the numerical study by Cessac [12] of the H\'enon map).  Both (i) and (ii) can occur only for dynamical systems that are not uniformly hyperbolic but, more specifically, they are related to ``energy nonconservation''.  This means that if the system is subjected to a periodic perturbation of small amplitude, the expectation value of observables may undergo a change of arbitrary large amplitude: the system is not passive or inert, it gives away energy to the outside world.  To be specific, we can say that the dynamical system $(f^t)$ with SRB measure $\rho$ is {\it active} if the corresponding susceptibility has singularities with ${\rm Im}\omega>0$.  Such singularities are expected in systems where "folding" causes tangencies between stable and unstable manifolds, as happens for the H\'enon map.  Indeed, the unimodal maps are a 1-dimensional model for folding in smooth dynamical systems.
\bigskip\noindent
{\bf 5. Conclusions.}
\medskip
	To understand linear response for a NESS away from equilibrium, one is led to investigating linear response for a general smooth dynamical system on a compact manifold.  It turns out that for hyperbolic dynamical systems, the theory of linear response is similar to the classical theory close to equilibrium: The Kramers-Kronig relations hold, and the fluctuation-dissipation theorem is modified by taking into account "oscillations of the system around its attractor".  For nonhyperbolic systems, linear response may fail in the sense that the NESS does not depend differentiably on the parameters of the system.  But this nondifferentiability may not be visible in physical situations.  Perhaps more important is the failure of dispersion relations: there may be singularities of the susceptibility in the upper half complex plane.  This happens when the system under consideration is {\it active}: physically this means that it can give away energy to the outside world.  Clearly, the "close to equilibrium" paradigm has to be drastically revised in the case of active systems.  It appears that both mathematical analysis and numerical simulations will be necessary to proceed to a new paradigm covering active systems, and to see in particular what happens to the Gallavotti-Cohen fluctuation theorem [19].
\bigskip\noindent
{\bf References.}
\medskip
[1] L. Andrey.  ``The rate of entropy change in non-Hamiltonian systems.''  Phys. Letters {\bf 11A},45-46(1985).

[2] V. Baladi.  {\it Positive transfer operators and decay of correlations.}  World Scientific, Singapore, 2000.

[3] V. Baladi and D. Smania.  "Linear response formula for piecewise expanding unimodal maps."  Nonlinearity {\bf 21},677-711(2008).

[4] V. Baladi and D. Smania.  Work in progress.

[5] P. Billingsley.  {\it Ergodic theory and information.}  John Wiley, New York, 1965.

[6] C. Bonatti, L. Diaz, and M. Viana.  {\it Dynamics beyond uniform hyperbolicity.} Springer, Berlin, 2005.

[7] R. Bowen.  ``Markov partitions for Axiom A diffeomorphisms.''  Amer. J. Math. {\bf 92},725-747(1970).

[8] R. Bowen.  ``Symbolic dynamics for hyperbolic flows.''  Amer. J. Math. {\bf 95},429-460(1973).

[9] R. Bowen.  {\it Equilibrium states and the ergodic theory of Anosov diffeomorphisms.} Lecture Notes in Math. {\bf 470}, Springer, Berlin,
1975.; 2nd ed, 2008.

[10] R. Bowen and D.Ruelle.  ``The ergodic theory of Axiom A flows.''  Invent. Math. {\bf 29},181-202(1975).

[11] O. Butterley and C. Liverani  "Smooth Anosov flows: correlation spectra and stability."  J. Modern Dynamics {\bf 1},301-322(2007).

[12] B. Cessac  "Does the complex susceptibility of the H\'enon map have a pole in the upper half plane?  A numerical investigation."  Nonlinearity {\bf 20},2883-2895(2007).

[13] N. Chernov.  ``Markov approximations and decay of correlations for Anosov flows.''  Annals of Math. {\bf 147},269-324(1998).

[14] D. Dolgopyat.  ``Decay of correlations in Anosov flows.''  Ann. of Math. {\bf 147},357-390(1998).

[15] D. Dolgopyat.  ``Prevalence of rapid mixing in hyperbolic flows.''  Ergod. Th. and Dynam. Syst. {\bf 18},1097-1114(1998).  ``Prevalence of rapid mixing-II: topological prevalence''  Ergod. Th. and Dynam. Syst. {\bf 20},1045-1059(2000).

[16] D. Dolgopyat.  "On differentiability of SRB states for partially hyperbolic systems"  Invent. Math. {\bf 155},389-449(2004).

[17] D.J. Evans and G.P. Morriss.  {\it Statistical mechanics of nonequilibrium fluids.}  Academic Press, New York, 1990.

[18] M. Field, I. Melbourne, A. T\"or\"ok.  ``Stability of mixing for hyperbolic flows.''  Ann. of Math. {\bf 166},269-291(2007).

[19] G. Gallavotti and E.G.D. Cohen.  ``Dynamical ensembles in stationary states.'' J. Statist. Phys. {\bf 80},931-970(1995).

[20] W.G. Hoover.  {\it Molecular dynamics.}  Lecture Notes in Physics {\bf 258}.  Springer, Heidelberg, 1986.

[21] K. Jacobs.  {\it Lecture notes on ergodic theory.}  Aarhus Matematisk Institut 1962/63.

[22] M.V. Jakobson.  "Absolutely continuous invariant measures for one-parameter families of one-dimensional maps."  Commun. Math. Phys. {\bf 81},39-88(1981).

[23] Y. Jiang and D. Ruelle  "Analyticity of the susceptibility function for unimodal Markovian maps of the interval."  Nonlinearity {\bf 18},2447-2453(2005).

[24] A. Katok, G. Knieper, M. Pollicott, and H. Weiss.  "Differentiability and analyticity of topological entropy for Anosov and geodesic flows."  Invent. Math. {\bf 98},581-597(1989).

[25] F. Ledrappier and J.-M. Strelcyn.  ``A proof of the estimation from below in Pesin's entropy formula.''  Ergod. Th. and Dynam. Syst. {\bf 2},203-219(1982).

[26] F. Ledrappier and L.S.Young.  ``The metric entropy of diffeomorphisms: I. Characterization of measures satisfying Pesin's formula.  II. Relations between entropy, exponents and dimension.''  Ann. of Math. {\bf 122},509-539,540-574(1985).

[27] C. Liverani.  ``On contact Anosov flows.''  Ann. of Math. {\bf 159},1275-1312(2004).

[28] V. Lucarini.  "Response theory for equilibrium and nonequilibrium statistical mechanics: causality and generalized Kramers-Kronig relations." J. Stat. Phys. {\bf 131},543-558(2008).

[29] V. Lucarini.  "Evidence of dispersion relations for the nonlinear response of the Lorenz 63 system."  arXiv:0809.0251v1

[30] M. Misiurewicz  "Absolutely continuous measures for certain maps of an interval."  Publ. Math. IHES {\bf 53},17-52(1981).

[31] V.I. Oseledec.  ``A multiplicative ergodic theorem.  Lyapunov characteristic numbers for dynamical systems.''  Tr. Mosk. Mat. Ob\v s\v c. {\bf 19},179-210(1968).  English translation, Trans. Moscow Math. Soc. {\bf 19},197-221(1968).

[32] W. Parry and M. Pollicott.  {\it Zeta functions and the periodic orbit structure of hyperbolic dynamics.}  Ast\'erisque {\bf 187-188}, Soc. Math. de France, Paris, 1990.

[33] Ya.B.Pesin.  ``Invariant manifold families which correspond to non-vanishing characteristic exponents.''  Izv. Akad. Nauk SSSR Ser. Mat. {\bf 40},No 6,1332-1379(1976).  English translation: Math. USSR Izv. {\bf 10},No 6,1261-1305(1976).

[34] Ya.B.Pesin.  ``Lyapunov characteristic exponents and smooth ergodic theory.''  Uspehi Mat. Nauk {\bf 32},No 4,55-112(1977).  English translation:  Russian Math. Surveys. {\bf 32},No 4,55-114(1977).

[35] D.Ruelle.  ``A measure associated with Axiom A attractors.''  Am. J. Math. {\bf 98},619-654(1976).

[36] D.Ruelle.  ``Ergodic theory of differentiable dynamical systems.''  Publ. Math. IHES {\bf 50},27-58(1979).

[37] D. Ruelle.  {\it Thermodynamic formalism.}  Addison-Wesley, Reading (Mass.),1978.  Second edition: Cambridge University Press, Cambridge, 2004.

[38] D. Ruelle.  "Differentiation of SRB states.''  Commun. Math. Phys. {\bf 187},227-241(1997); ``Correction and complements.''  Commun. Math. Phys. {\bf 234},185-190(2003).

[39] D. Ruelle.  ``General linear response formula in statistical mechanics, and the fluctuation-dissipation theorem far from equilibrium.''  Phys. Letters {\bf A 245},220-224\ (1998).

[40] D. Ruelle.  ``Application of hyperbolic dynamics to physics: some problems and conjectures.''  Bull. Amer. Math. Soc. (N.S.) {\bf 41},275-278(2004).

[41] D. Ruelle.  "Differentiating the absolutely continuous invariant measure of an interval map f with respect to f."  Commun.Math. Phys. {\bf 258},445-453(2005).

[42] D. Ruelle.  "Differentiation of SRB states for hyperbolic flows."  Ergod. Theor. Dynam. Syst. {\bf 28},613-631(2008).

[43] D. Ruelle.  "Structure and $f$-dependence of the a.c.i.m. for a unimodal map $f$ of Misiurewicz type."  Commun. Math. Phys., to appear.

[44] Ya.G. Sinai.  ``Markov partitions and C-diffeomorphisms.''  Funkts. Analiz i Ego Pril. {\bf 2}, No {\bf 1},64-89(1968).  English translation, Functional Anal. Appl. {\bf 2},61-82(1968).

[45] Ya.G. Sinai.  ``Constuction of Markov partitions.''  Funkts. Analiz i Ego Pril. {\bf 2}, No {\bf 3},70-80(1968).  English translation, Functional Anal. Appl. {\bf 2},245-253(1968).

[46] Ya.G. Sinai.  ``Gibbsian measures in ergodic theory.''  Uspehi Mat. Nauk {\bf 27}, No {\bf 4},21-64(1972).  English translation, Russian Math. Surveys {\bf 27}, No {\bf 4},21-69(1972).

[47] S. Smale.  ``Differentiable dynamical systems.''  Bull. AMS {\bf 73},747-817(1967).

[48] H. Spohn and J.L. Lebowitz.  ``Stationary non-equilibrium states of infinite harmonic systems.''  Commun. Math. Phys. {\bf 54},97-120(1977).

[49] J. Toll.  "Causality and the dispersion relation: logical foundations." Phys. Rev. {\bf 104},1760-1770(1956).

[50] H. Whitney.  ``Analytic expansions of differentiable functions defined in closed sets.''  Trans. Amer. Math. Soc. {\bf 36},63-89(1934).

[51] L.-S. Young.  ``What are SRB measures, and which dynamical systems have them?''  J. Statist. Phys. {\bf 108},733-754(2002).

\end